# Investigation of Fe-Ag and Ag-Fe Interfaces in Ag-57Fe-Ag trilayer Using Nuclear Resonance Scattering under X-ray Standing Wave Conditions

Manisha Priyadarsini[1], Sharanjeet Singh[1], Ilya Sergeev[2], Olaf Leupold[2], Ajay Gupta[3], Ilaria Carlomagno[4], V Raghavendra Reddy[1], K. R. Priolkar[1], Dileep Kumar[1,*]

[1]*UGC-DAE Consortium for Scientific Research, Khandwa Road, Indore -452001, India*
[2]*Deutsches Elektronen-Synchrotron DESY, Hamburg 22607, Germany*
[3]*Department of Physics, University of Petroleum and Energy Studies, Bidholi Campus, Dehradun 248007, India*
[4]*Elettra Sincrotrone Trieste, Strada Statale 14, km 163.5 in AREA Science Park, Basovizza, Trieste 34149, Italy*
[5]*Raja Ramanna Centre for Advanced Technology, Indore-452013, India*

## Abstract

Interfaces in magnetic multilayers play a decisive role in determining their structural and magnetic properties, which significantly impact the performance of the spintronic devices. This work investigates the bottom and top interfaces of a few-nanometer-thick $^{57}$Fe layer in the Ag/$^{57}$Fe/Ag trilayers. The sub-nanometer depth resolution was obtained using X-ray standing waves (XSWs) generated by an underlying $[W/Si]_{10}$ multilayer. Each interface was selectively probed by scanning the standing-wave antinodes across the trilayer by varying the X-ray incidence angle around the Bragg condition. Detailed depth-resolved structural and magnetic information was obtained by combining the X-ray fluorescence (XRF) with isotope-selective ($^{57}$Fe) grazing-incidence nuclear resonance scattering (GINRS) under XSW conditions. The results reveal a pronounced asymmetry in the interfacial roughness; the $^{57}$Fe-on-Ag interface exhibits a larger r.m.s. roughness of 9.7 ± 1.0 Å, whereas the Ag-on-$^{57}$Fe interface is comparatively smoother, with a roughness of 6.4 ± 1.0 Å. Correspondingly, GINRS measurements probed distinct hyperfine fields at the two interfaces, consistent with the observed structural asymmetry. Temperature-dependent measurements reveal enhanced interfacial intermixing. Fe K-edge extended X-ray absorption spectroscopy (EXAFS) combined with temperature-dependent magnetic measurements shows that annealing at 325 °C induces strong nanoscale intermixing, resulting in the formation of Fe nanoparticles embedded in Ag. These results demonstrate that XSW-assisted XRF and GINRS provide a powerful and interface-specific probe of buried structural and magnetic asymmetries, offering valuable insights for the design and optimization of spintronic multilayer nanostructures.

# Introduction

Magnetic multilayers (MLTs) are key components of modern spintronic technologies, including magnetoresistive random-access memory (MRAM), magnetic field sensors, racetrack memories, and emerging skyrmion-based devices [1]. In these systems, information is stored and processed using the electron spin rather than its charge. The magnetic behaviour is controlled by interlayer exchange coupling, magnetic anisotropy, and spin-dependent scattering, all of which are strongly affected by buried interfaces between different layers [2–4]. As a result, device performance, such as switching speed, energy efficiency, and stability, depends not only on the bulk properties of the materials but, to a large extent, on the atomic-scale structure and chemistry of these interfaces. Even small variations in interface roughness, atomic intermixing, or layer continuity can significantly modify magnetic anisotropy [5,6], spin transport, and the motion of domain walls or skyrmions [7,8]. Such changes may lead to slower switching, enhanced pinning, or reduced magnetic stability [9]. In thin-film growth, the interface formed when material X is deposited on material Y may differ from the inverse interface formed when Y is deposited on X. This is due to the differences in surface free energy, atomic mobility, and chemical reactivity during growth [10]. Consequently, the two interfaces of a single magnetic layer often exhibit different roughness and interdiffusion [11]. Atomic-scale diffusion and structural asymmetry at buried interfaces have therefore emerged as important factors governing magnetic properties in MLT spintronic systems.

A variety of experimental techniques have been used to study the structural properties of buried interfaces, including X-ray reflectivity (XRR) [12,13], *in situ* scanning tunneling microscopy (STM) [14], Rutherford backscattering (RBS) [15], and Auger electron spectroscopy [16,17]. While these methods provide valuable information, they suffer from limitations when applied to precise and true buried interfaces. Surface-sensitive techniques probe only the topmost layers and cannot access the buried interfaces. Depth-sensitive techniques often provide signals averaged over the entire MLT and therefore lack the spatial resolution required to distinguish the two inequivalent interfaces of a single trilayer. Furthermore, magnetic MLTs are commonly studied as stacks of repeated bilayers [15,18], in which the measured signals average over many interfaces. This averaging masks interface-specific structural and magnetic contributions.

Fe/Ag heterostructures provide a model system for addressing these challenges. Fe and Ag are immiscible and have a very large difference between their surface free energies, which may lead to pronounced interface asymmetry depending on the deposition sequence. In addition, the weak spin-orbit coupling of Ag minimizes relativistic effects, allowing the role of structural and chemical asymmetry to be studied more directly. The Fe/Ag interface plays an important role in phenomena such as interlayer exchange coupling, giant magnetoresistance, perpendicular magnetic anisotropy [19], and spin transport [20–22]. Despite extensive studies using conventional techniques, the relationship between interface-specific structure and magnetic properties in this system remains unclear [23,24]. To overcome these limitations, it is essential to probe both inequivalent interfaces independently within the same sample and under identical conditions.

In this work, we investigate an Ag/$^{57}$Fe/Ag trilayer using synchrotron-based, isotope-sensitive grazing-incidence nuclear resonance scattering (GI-NRS) combined with X-ray standing-wave (XSW) techniques. The use of the $^{57}$Fe isotope provides element- and isotope-specific magnetic sensitivity [10,12,25–29]. When the X-rays fall on the sample at an angle around the Bragg peak of the [W/Si] MLT, standing waves are generated in the sample, which extend beyond the [W/Si] MLT and into the Ag /Fe/Ag trilayer. Elemental concentration depth profiles are obtained by monitoring Fe X-ray fluorescence (XRF) as the antinode is swept across the interfaces. This also allows the precise angles at which an antinode overlaps an individual interface to be studied using GINRS for the interface-resolved magnetic properties of both the Ag-on-$^{57}$Fe and $^{57}$Fe-on-Ag interfaces within the same trilayer [30]. Using this approach, we directly measure the subtle differences in the magnitude and orientation of the hyperfine field at the two interfaces. Finally, extended X-ray absorption fine structure (EXAFS) spectroscopy was performed as an element-specific probe of the local atomic structure, yielding detailed information on interatomic distances, coordination numbers, and the types of neighboring atoms. These results highlight the importance of interface asymmetry in determining magnetic behaviour and demonstrate a powerful, non-destructive method for interface-resolved studies of buried magnetic heterostructures. The findings are relevant to a broad range of magnetic MLT systems used in spin-valves, tunnelling magnetoresistance devices, and high-density magnetic storage technologies.

## Experimental Details

The investigated sample consists of an Ag/$^{57}$Fe/Ag trilayer deposited on a periodic $[W/Si]_{10}$ multilayer structure, designed to generate XSW for depth-resolved interface studies. The complete structure is [W (20 Å)/Si (30 Å)]$_{\times 10}$/Si (9Å)/Ag (25 Å)/$^{57}$Fe (40 Å)/Ag (25 Å)/Si (120 Å) . This nominal-thickness layer was deposited by ion-beam sputtering using a Kaufman-type 3 cm-wide beam ion source in a vacuum chamber maintained at a base pressure of ~$2 \times 10^{-8}$ mbar. Thicknesses are obtained using a calibrated thickness monitor in the deposition chamber. All targets had a purity of 99.9%, with the iron target enriched to 95% of $^{57}$Fe. $[W/Si]_{10}$ was deposited to generate the x-ray standing wave, whereas Si layer 9Å is deposited between $[W/Si]_{10}$ and the Ag/Fe/Ag trilayer structure, so as to optimize overlap of XSW antinodes with the two interfaces of the $^{57}$Fe layer ($^{57}$Fe-on -Ag and Ag-on-$^{57}$Fe). It is required to obtain the interface-resolved information under XSW conditions. The layer thicknesses are finalized based on a separate sample, prepared under identical conditions with all layer thicknesses held constant, except for the Si layer, which is wedge-shaped (thickness ranging from 3 to 20 Å). The wedge-shaped Si layer deposited on top of the $[W/Si]_{10}$ multilayer allows continuous tuning of the vertical position of the Ag/$^{57}$Fe/Ag trilayer with respect to the XSW field generated by the underlying mirror. XRR and XRF measurements were performed to determine the structural parameters of the multilayer and the depth distribution of $^{57}$Fe. XRR measurements provide structural information dominated by the $[W/Si]_{10}$ multilayer, which is essential for accurate XSW field calculations. In contrast, XRF measurements are element-specific and therefore directly probe the Fe layer and its interfaces. XRR and XRF measurements were carried out using a Bruker D8 diffractometer with a beam energy of 8.047 keV. XRF spectra were recorded using an Amptek XR-100T/CR PIN diode detector with an energy resolution of approximately 250 eV. Simultaneous fitting of XRR (using Parratt's formalism) [34] and XRF data was performed, enabling the extraction of interface roughness and depth-resolved $^{57}$Fe density profiles with sub-nanometer resolution. Magnetic properties were investigated using conversion electron Mössbauer spectroscopy (CEMS), GI-NRS, and magneto-optic Kerr effect (MOKE) measurements. The CEMS measurements were performed using a Wissel Mössbauer spectrometer and a gas flow proportional counter with a 95% He and 5% $CH_4$ mixture. GI-NRS measurements were carried out at the P01 beamline of PETRA III (DESY, Hamburg) using 14.4 keV synchrotron radiation with a beam size of 200 × 200 μm$^2$ (h × v). The synchrotron was operated in 40-bunch mode with a bunch separation of 192 ns for GI-NRS measurements. An avalanche photodiode detector was used for GI-NRS measurement, having a time resolution of ∼1 ns. Under X-ray standing-wave conditions, GI-NRS enables depth-selective probing of the $^{57}$Fe layer [35]. The MOKE microscope (M/s Evico Magnetics, Germany) equipped with a broadband white LED light source (with incident angle ∼ 45°) was utilized to obtain the hysteresis loops. The measurements were performed in the longitudinal MOKE geometry over a magnetic field range of ±300 Oe, which is appropriate for probing the in-plane magnetic response of the sample. To investigate the thermal stability of the interfaces and the evolution of magnetic properties, isochronal annealing was performed at 225 °C and 325 °C for 1 hour under a vacuum of ~5 × $10^{-7}$ mbar. The lower annealing temperature primarily probes early-stage interfacial intermixing, while the higher temperature is intended to drive extensive Fe-Ag interdiffusion, leading to strong magnetic dilution and the loss of long-range ferromagnetic order. After each annealing step, the annealed sample was subsequently characterized using XRR, XRF, CEMS, GI-NRS, and MOKE measurements. Local atomic structure around Fe atoms was investigated using Fe K-edge (E = 7112 eV) extended X-ray absorption fine structure (EXAFS) measurements with a beam size of 0.1 × 0.5 mm$^2$ (h × v). EXAFS experiments were performed in fluorescence geometry in the vacuum chamber available at XRF beamline 10.1L of ELETTRA (Trieste, Italy) at a grazing-incidence angle corresponding to the XRF maximum, thereby enhancing sensitivity to the $^{57}$Fe-containing region [36,37]. Standard data reduction and fitting were carried out using the ATHENA and ARTEMIS software packages [38]. Scanning electron microscopy (SEM) combined with energy-dispersive X-ray spectroscopy (EDS) was used as a complementary probe to assess large-scale morphological and compositional uniformity of the films.

# Theoretical Background

XSW techniques are now becoming a crucial tool for depth-resolved measurements in a variety of systems, including studies of metal/metal and magnetic MLT, magnetism in buried layers, atomic migration, interface intermixing, and annealing-induced structural changes [39–42]. In this study, the theoretical framework of XSW is employed to probe the interfacial asymmetries in an Ag/$^{57}$Fe/Ag trilayer system. The XSW is generated within the MLT structure by the coherent scattering of monochromatic X-rays, resulting in a spatial modulation of electromagnetic field intensity, which allows depth-selective excitation of resonant nuclei [28]. If we consider a MLT structure, the intensity of the secondary radiation ($A_f\ (z)$) created at depth *z* is proportional to the square modulus of the total X-ray field amplitude *E (z)* at depth *z* and is given by,

$$A_f(z) \sim N_F(z)|E(z)|^2 = N_F(z)|E^t(z) + E^r(z)|^2 \qquad (1)$$

Where $E^t(z)$ and $E^r(z)$ are the amplitudes of incident and reflected waves, $N_F(z)$ is the volume density of the fluorescent atoms. The positions of the standing wave antinodes can be shifted by slightly varying the incidence angle of X-rays [27,35,39]. The reflectance of an ultrathin layer positioned *z* away from a MLT, $I_R$, in the simplest scenario, is determined by

$$I_R = |r\ E(z)^2|^2 \qquad (2)$$

where *r* is the reflectivity amplitude from the same layer in vacuum. It follows from (2) that the reflectivity from an ultrathin layer is enhanced by the fourth power of the standing wave amplitude *E(z)*, and that is true in the general case. Therefore, standing waves have an even greater impact on reflectivity than on secondary radiation. The problem of how to extract the contribution from a selected layer in the total reflectivity intensity can be resolved by the resonant energy or time reflectivity dependence. So resonant reflectivity has a great potential for depth-selective investigations.

Nuclear resonant reflectivity from the films containing Mössbauer isotopes is normally measured with synchrotron radiation in the time domain. Reflectivity amplitudes in the energy *R (ω)* and time *R (t)* domains are connected by the Fourier transform.

$$R(t) = \frac{1}{2\pi}\int_{-\infty}^{\infty} R(\omega)e^{-i\omega t}d\omega \qquad (3)$$

So, for each grazing angle, θ the time spectrum of the nuclear resonant reflectivity can be calculated,

$$I_{nucl}(t,\theta) = A\ |R(t,\theta)|^2 \qquad (4)$$

A normalization factor *A* is determined by the total amount of resonant quanta in the bunches of the synchrotron radiation. When equation (4) is integrated over the delay time for each, the delayed nuclear resonant reflectivity curve is obtained:

$$I_{nucl}(\theta) = A\int_{-\infty}^{\infty}|R(t,\theta)|^2\ dt \qquad (5)$$

This equation characterizes the depth position of the resonant nuclei. It may be noted that the nuclear resonance part of the signal, which is proportional to the fourth power of the standing wave amplitude, offers significantly greater sensitivity than other XSW-based techniques, such as XRF. This enhancement forms the basis of our approach to distinguish signals originating from individual interfaces. The enhanced sensitivity is critical in resolving the interfacial asymmetries and magnetic transitions observed in this study.

## Results and Discussion

### (i) Structural and magnetic characterizations of the pristine sample:

The MLT structure used to study the interfaces in Ag/$^{57}$Fe/Ag trilayer is depicted in Fig. 1(a). The thicknesses shown in the figure are obtained after fitting the XRR and XRF data. The sample's XRR and the corresponding Scattering length density (SLD) versus thicknesses are shown in Fig. 1(b) and (c), respectively. The XRR profile of the MLT, measured in its as-prepared state, presents the reflected intensity as a function of the scattering vector, $q = (4\pi \sin\theta)/\lambda$, where $\theta$ is the angle of incidence and $\lambda$ is the wavelength of the incident X-ray. The distinct Bragg peak in the reflectivity indicates the bilayer periodicity of the underlying $[W/Si]_{10}$ MLT.

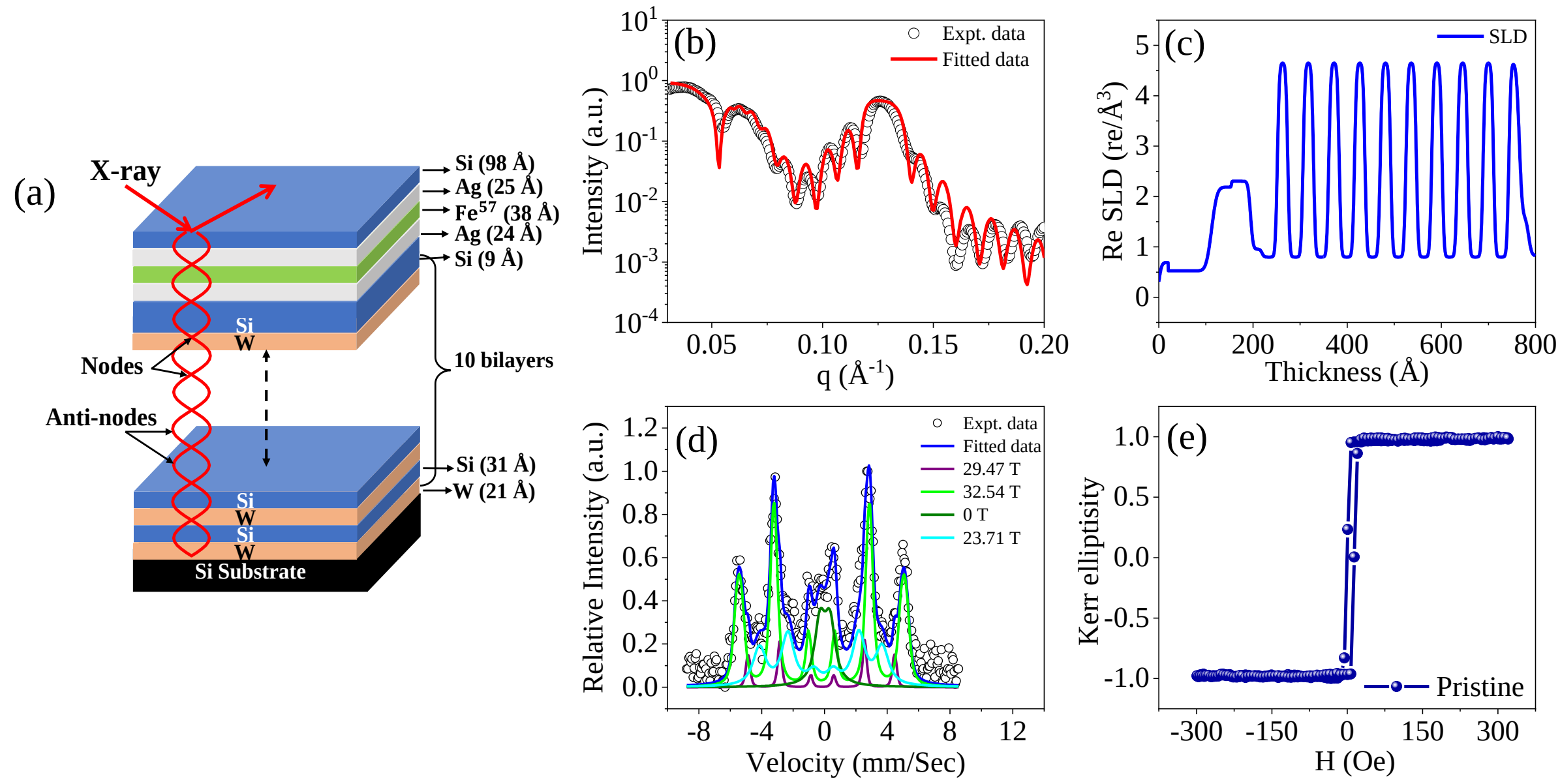

**Figure 1.** (a) Schematic of the multilayer structure. (b) The XRR plot and (c) the corresponding electron scattering length density profile for the as-prepared sample. The red lines represent the fitting to the experimental data. (d) depicts the CEMS spectra of the sample along with fitting, and (e) represents the hysteresis loop of the sample in the as-prepared state.

The XRR of the structure is fitted by using the GenX software [43]. It may be noted that the best fit to the data is obtained by considering the sample's thickness, density, and roughness as fitting parameters. The total film thickness can be obtained from the period of the Kiessig oscillations, whereas the bilayer periodicity and the average interface roughness of the underlying $[W/Si]_{10}$ MLT can be obtained from the position and intensity of the Bragg peak. Figure 1(d) displays the CEMS spectra of the pristine sample exhibiting a sextet structure. It probes the Bhf distribution averaged over the entire $^{57}Fe$-containing layer. The CEMS spectra of the as-prepared sample were fitted using the REFTIM program to find the contributions of the hyperfine field (Bhfs) distribution. The fitting resolves multiple Bhf components: a dominant bulk-like contribution around ~32 T, intermediate-field components in the range of ~29-23 T associated with partially disordered Fe environments, and a zero-field (~ 0 T) component attributed to magnetically diluted or weakly magnetic regions. Figure 1(e) shows the MOKE hysteresis loop of the pristine sample. It can be seen that the loop is square-shaped, with coercivity (Hc) ~ 7 ± 1 Oe and remanence (Mr/Ms) ~ 0.80 ±0.04.

It is important to note that the investigated sample comprises a complex multilayer architecture consisting of a $[W/Si]_{10}$ multilayer mirror, a Si spacer layer, an Ag/Fe/Ag trilayer, and a Si capping layer. Quantitative fitting of the XRR data for such a structure involves a large number of strongly correlated parameters, including the thickness, roughness, and electron density of each individual layer. Consequently, the XRR signal is predominantly governed by the underlying $[W/Si]_{10}$ multilayer mirror, limiting the sensitivity of XRR alone to the structural details of the Ag/Fe/Ag trilayer. In contrast, the simultaneous analysis of XRR and XRF data under X-ray standing-wave conditions provides enhanced depth selectivity and yields reliable structural information on the Ag/Fe/Ag trilayer, which is the primary focus of the present study [11].

Figure 2(a) gives the normalized $^{57}Fe$ fluorescence data recorded by varying the incident angle from 0.09 to 0.16 $Å^{-1}$ (across the XRR Bragg peak). To obtain precise quantitative information about the $^{57}Fe$ layer, Fe-XRF and XRR data are fitted simultaneously and are shown as a continuous line in Fig. 2(a). The $^{57}Fe$ fluorescence intensity as a function of q around the first Bragg peak has been analyzed for the sample. Two peaks (P1 and P2) were observed around the Bragg peak at q values corresponding to 0.121 and 0.135 $Å^{-1}$, respectively. From the fitting, the sample structure is obtained as [W (21 Å)/Si (31 Å)]$_{\times 10}$/Si (9 Å)/Ag (24 Å)/$^{57}Fe$ (38 Å)/Ag (25 Å)/Si (98 Å), and these thicknesses are also shown in the sample structure in Fig. 1(a). The best fit to the XRF data for the as-prepared sample yields root-mean-square (r.m.s.) roughness values of 6.4 ± 1.0 Å and 9.7 ± 1.0 Å for the top and bottom interfaces, respectively. These results indicate that the top interface is smoother than the bottom interface. To assess the reliability of the roughness, the fluorescence data were simulated with equal roughnesses (10 Å) at both interfaces and shown as the dotted line in Fig. 2(a). The normalized density profile of the $^{57}Fe$ layer obtained from the XRF fitting is shown in Fig. 2(b), which also shows two different density profiles at the two interfaces.

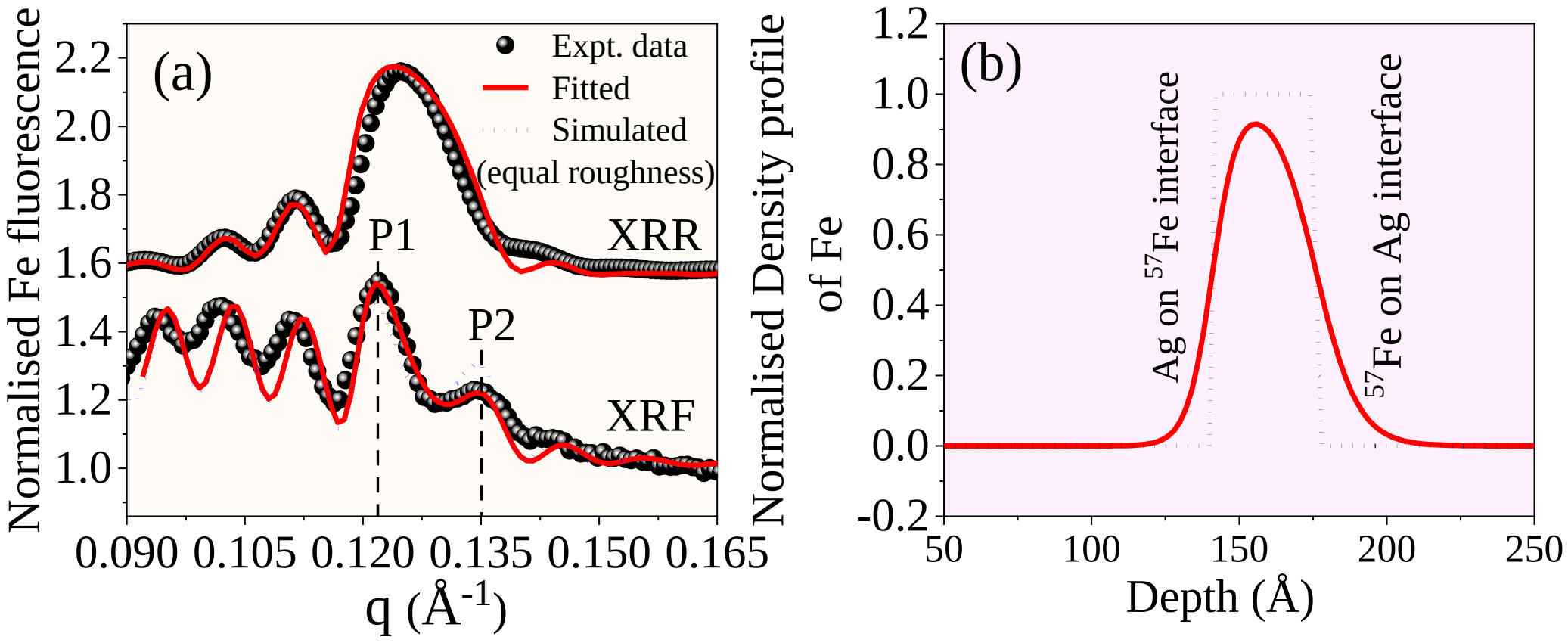

**Figure 2.** (a) shows the simultaneously fitted XRR and Fe XRF data of the pristine sample. The black spheres represent the experimental data, the red continuous line represents the simulated curve, and the dotted line shows the simulated data for equal roughness considered at both interfaces. (b) The normalized density profile of $^{57}$Fe as a function of depth.

Based on the simultaneously fitting Fe-XRF and XRR parameters of the sample structure, a contour plot was simulated to visualize the X-ray field intensity (XFI) distribution due to the formation of XSWs and the origin of the two distinct peaks in $^{57}$Fe fluorescence data from the single $^{57}$Fe layer. Fig. 3(a) presents the formation of XSW nodes and antinodes inside the Ag/$^{57}$Fe/Ag trilayer as a function of q. The position of the $^{57}$Fe layer is marked by shaded regions in this plot.

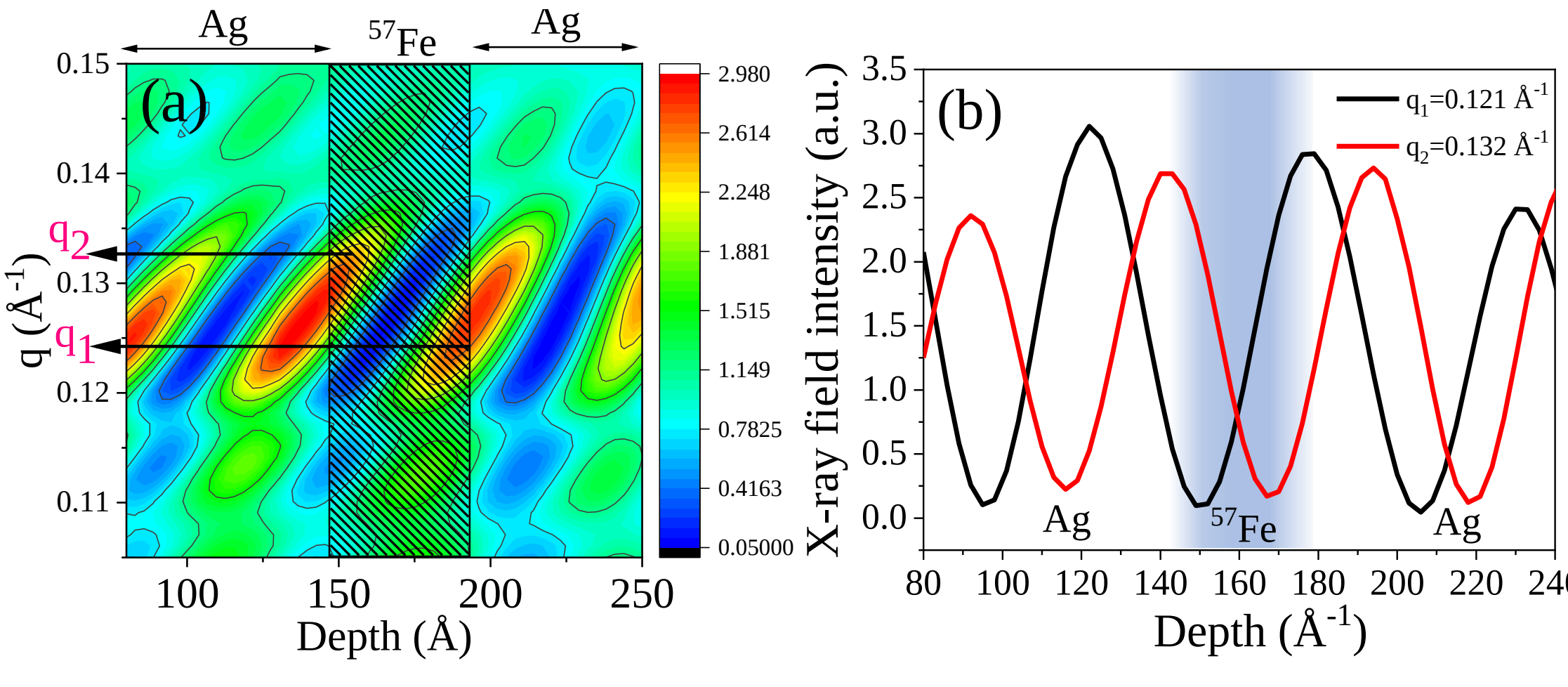

**Figure 3.** (a) Calculated X-ray field intensity profile produced by the $[W/Si]_{10}$ multilayer structure. The colour code represents the normalized X-ray field intensity (XFI). (b) represents the variation of standing wave field intensity along the $^{57}$Fe layer at $q_1$ = 0.121 Å$^{-1}$ ($TE_1$ mode), $q_2$ = 0.132 Å$^{-1}$ ($TE_2$ mode). The XFI is shown in arbitrary units (a.u.), normalized to the incident beam intensity.

It is clear that the positions of the antinodes varied across the $^{57}$Fe layer by adjusting the incident angle across the MLT Bragg Peak. The antinode at $q_2$ ~ 0.132 Å$^{-1}$ crosses through the top interface (i.e., the Ag-on-$^{57}$Fe) layer, whereas the bottom interface (i.e., $^{57}$Fe-on-Ag) overlaps with the antinode at $q_1$ ~

0.121 Å$^{-1}$. Figure 3(b) shows the XSW field intensity variations at the two interfaces corresponding to these two distinct q values. For $q_1$ = 0.121 Å$^{-1}$, the $r^{th}$ antinode partly overlaps with the bottom $^{57}$Fe interface, giving rise to the high-intensity peak in the fluorescence. As q increases, the antinode moves out of the $^{57}$Fe layer, thereby decreasing the fluorescence intensity. However, as q increases, the $(r+1)^{th}$ antinode moves inside and partially overlaps with the top interface, giving rise to the next peak at q = 0.132 Å$^{-1}$.

These conditions, characterized by overlapping antinodes at the two interfaces at distinct q values, enable depth-selective measurements with enhanced interface sensitivity. By performing X-ray-based experiments at these two q values, detailed information from both interfaces can be selectively extracted. In the present work, the interfacial magnetic properties were investigated using synchrotron-radiation-based nuclear resonance reflectivity (NRR) and grazing-incidence nuclear resonant scattering (GINRS) at these selected q values. GINRS is the time-domain analogue of isotope-sensitive Mössbauer spectroscopy and is governed by the nuclear transition energies of $^{57}$Fe in the sample [44,45]. Consequently, the nuclear resonant scattering amplitude depends sensitively on the magnetic hyperfine field and the spin orientation of $^{57}$Fe nuclei, giving rise to characteristic quantum beat patterns in the GINRS time spectra [44–46]. These quantum beats provide direct information on the magnetic hyperfine structure and the orientation of magnetization in thin-film systems.

Figure 4(a) shows the schematic for the GINRS measurements. Here, $\alpha_i$ and $\alpha_f$ stand for the incidence and reflected angles of the synchrotron beam, while $k_i$ and $k_f$ represent the incident and reflected wave vectors. Before GINRS-time spectra measurements, XRR and NRR from the $^{57}$Fe layer were measured at 14.4 keV with increasing incident angles, similar to XRF measurements, to select the appropriate angle for interface-resolved GINRS (nuclear resonance scattering vs time) measurements. NRR is the reflectivity of the resonant counts with increasing angle; it is only sensitive to the $^{57}$Fe. This is in contrast to XRR, where the reflected counts are due to electronic scattering and are therefore sensitive to all layers present in the samples [12,13]. In light of these facts, a combined analysis of XRR and NRR, along with GI-NRS data, can provide detailed information on magnetism and structure at the interfaces.

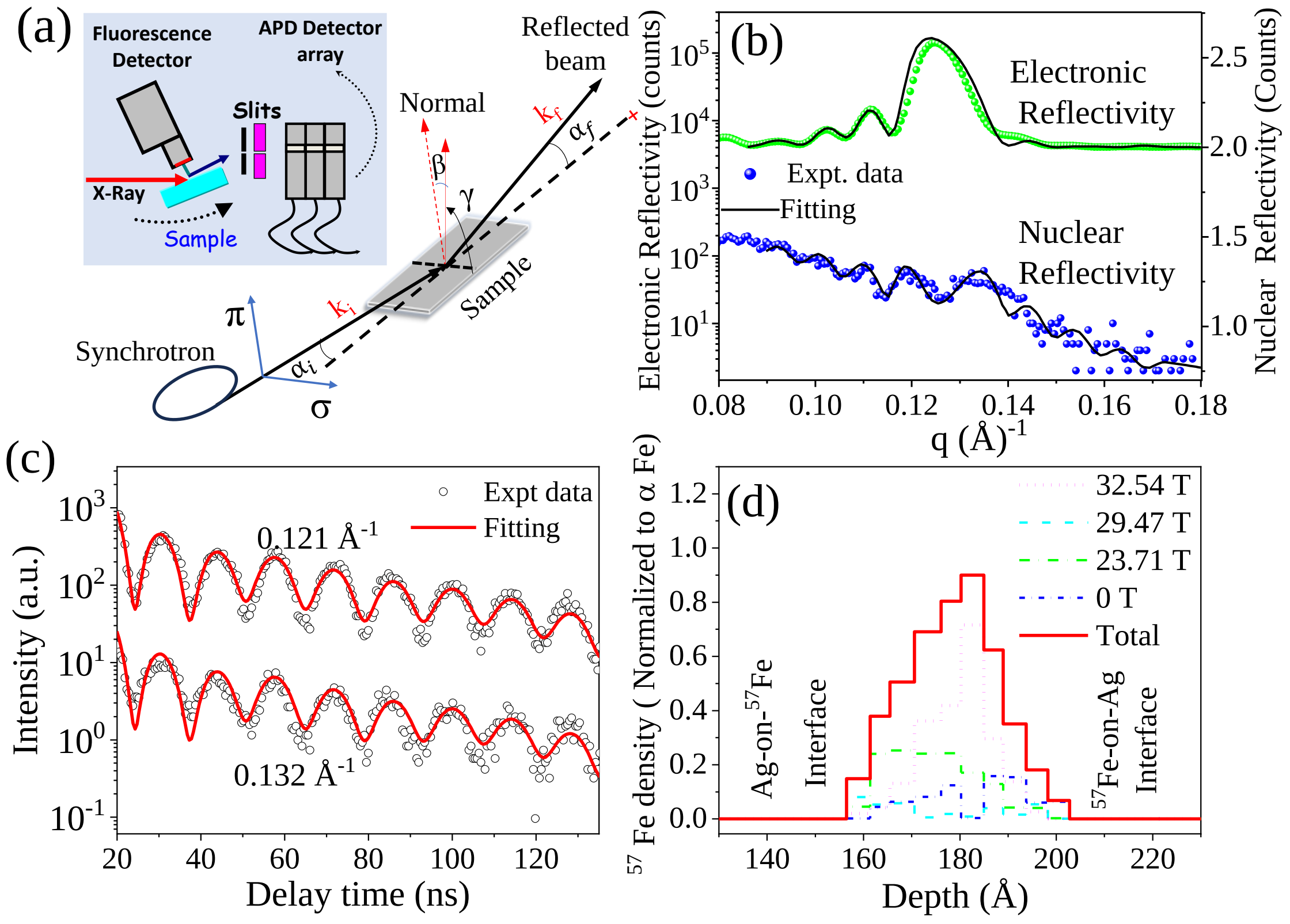

**Figure 4.** (a) Shows the schematic of the GINRS measurements. (b) Represents the electronic and the nuclear reflectivity, and (c) denotes the GI-NRS time spectra of the sample recorded at two incident angles (0.121 and 0.132 Å$^{-1}$). The intensity is presented in linear scale in (b) and log scale in (c). (d) denotes the depth distribution of $^{57}$Fe nuclei, with contributions from different hyperfine fields, obtained from the simultaneous fitting of nuclear reflectivity and time spectra.

The combined plot of the fitted XRR and NRR is shown in Fig. 4(b). The reflectivity data in Fig. 4 (b) are shown in the Bragg-angle region, where the NRR peak appears close to the multilayer Bragg reflection. As compared to one XRR Bragg peak at q=0.128 Å$^{-1}$, two peaks at $q_1$=0.121 Å$^{-1}$and $q_2$=0.132 Å$^{-1}$ in NRR are due to the overlap of antinodes with the bottom and top interface side of the $^{57}$Fe layer. The origin of the two peaks in the NRR is similar to the XRF (previous section), where the antinode crosses the $^{57}$Fe interfaces at two distinct q values. But, the different relative peak intensity in NRR compared to XRF is mainly due to the different scattering dependence of resonance data on the standing wave amplitude [45]. To further obtain interface-weighted magnetic information from the $^{57}$Fe layer, GINRS measurements were performed, and GINRS spectra (time spectra) were collected (Fig. 4 (c)) at two incident angles (0.121 Å$^{-1}$ and 0.132 Å$^{-1}$), where the antinode overlaps with the bottom and the top interfaces. The GI-NRS time spectra were analyzed in REFTIM using hyperfine-field components ( 32.54 T, 29.47 T, 23.71 T, and ~ 0 T) derived from the CEMS fit in the previous section. Unlike CEMS, GI-NRS time spectra were fitted by taking different relative fractions of these components along the depth of the $^{57}$Fe layer, and fitting was refined iteratively until satisfactory agreement with the experimental time spectra was achieved [47]. The depth-dependent distribution of the fitted Bhf components corresponding to 32.54 T, 29.47 T, 23.71 T, and ~ 0T, along with their relative fractions, is shown in Fig. 4(d). It summarizes the relative contributions of different Bhf components across the entire $^{57}$Fe layer.

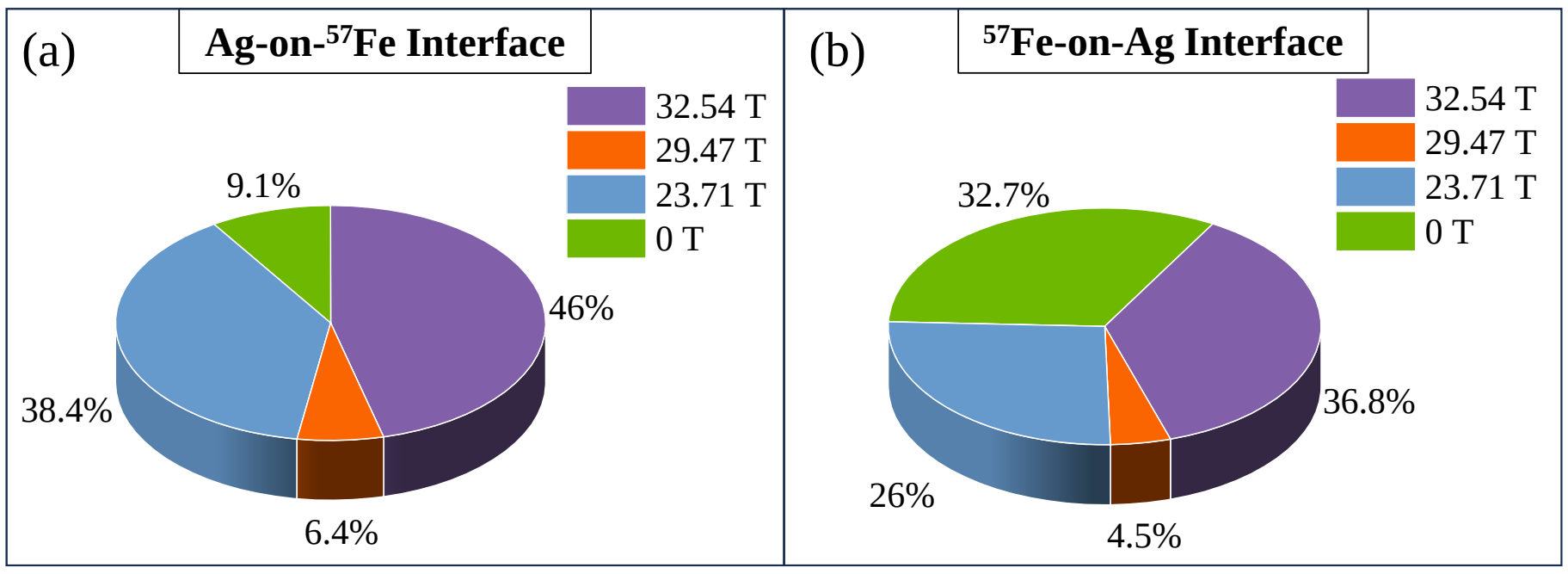

**Figure 5.** Percentage contribution of Bhfs in samples at (a) Ag-on-$^{57}$Fe and (b) $^{57}$Fe-on-Ag interface.

The density distributions of the Bhfs at the top (Ag-on-$^{57}$Fe) and bottom ($^{57}$Fe-on-Ag) interfaces are also extracted and resolved separately in pie charts shown in Fig. 5. This presents the fitting results in a more compact form, illustrating the relative fractions of Bhf components at both interfaces. It is clear that the top interface is dominated by bulk-like Fe environments, with the highest-field component (32.54 T) contributing 46%, compared to 36.8% at the bottom interface. Similarly, the intermediate-field components at 29.47 T and 23.71 T account for 6.4% and 38.4% at the top interface, respectively, while their contributions at the bottom interface are reduced to 4.5% and 26%. In contrast, the bottom interface exhibits a substantially larger fraction of low Bhfs. The zero-field component (~ 0 T), associated with strongly diluted or superparamagnetic-like Fe environments, contributes 32.7% at the bottom interface, which is significantly higher than the 9.1% observed at the top interface. This enhanced low-field weight at the bottom interface indicates increased Ag coordination and reduced $^{57}$Fe-$^{57}$Fe exchange interactions. Overall, the top interface retains a higher proportion of bulk-like Fe sites, consistent with a more ordered local environment and stronger magnetic coupling, whereas the bottom interface shows a broader Bhf distribution with a dominant low-field contribution. These results clearly demonstrate a magnetic asymmetry between the two interfaces, directly reflecting their structurally inequivalent nature.

### (ii) Effect of thermal annealing:

Thermal annealing was carried out to investigate the evolution of the interfaces under XSW conditions and the associated changes in magnetic properties in the Ag/$^{57}$Fe/Ag trilayer. The W/Si multilayer mirror may influence the initial nucleation and growth of the bottom Ag layer and thereby affect the as-deposited microstructure of the Ag/$^{57}$Fe/Ag trilayer. Similar substrate- and underlayer-dependent effects have been reported in Fe/Ag multilayer systems, where the stacking sequence can influence the grain size and magnetic properties of Fe layers [Kiss et al., Phys. Rev. B 98, 144423 (2018)]. However, in the present study, all annealing experiments were performed on the same sample architecture. Therefore, any influence of the W/Si multilayer remains constant and is incorporated into the initial state of the sample. XRR and XRF patterns obtained after annealing at 225 °C and 325 °C, are presented in Fig. 6(a-d). It may be noted that XRR patterns in Fig. 6(a) and (b), the Bragg peak position and intensity remain unchanged even after annealing at 325 °C, indicating that the $[W/Si]_{10}$ MLT remains structurally stable up to 325 °C. In contrast, the X-ray fluorescence (XRF) signal collected in the angular range encompassing the Bragg peak exhibits pronounced changes upon annealing. After annealing at 225 °C, the two XRF peaks [Fig. 2(a)] merge into a single peak centered at a lower momentum transfer, q = 0.125 $Å^{-1}$. Upon further annealing at 325 °C, the XRF response evolves into a single broad peak centered at q = 0.127 $Å^{-1}$, with its peak position coinciding with the XRR Bragg peak maximum.

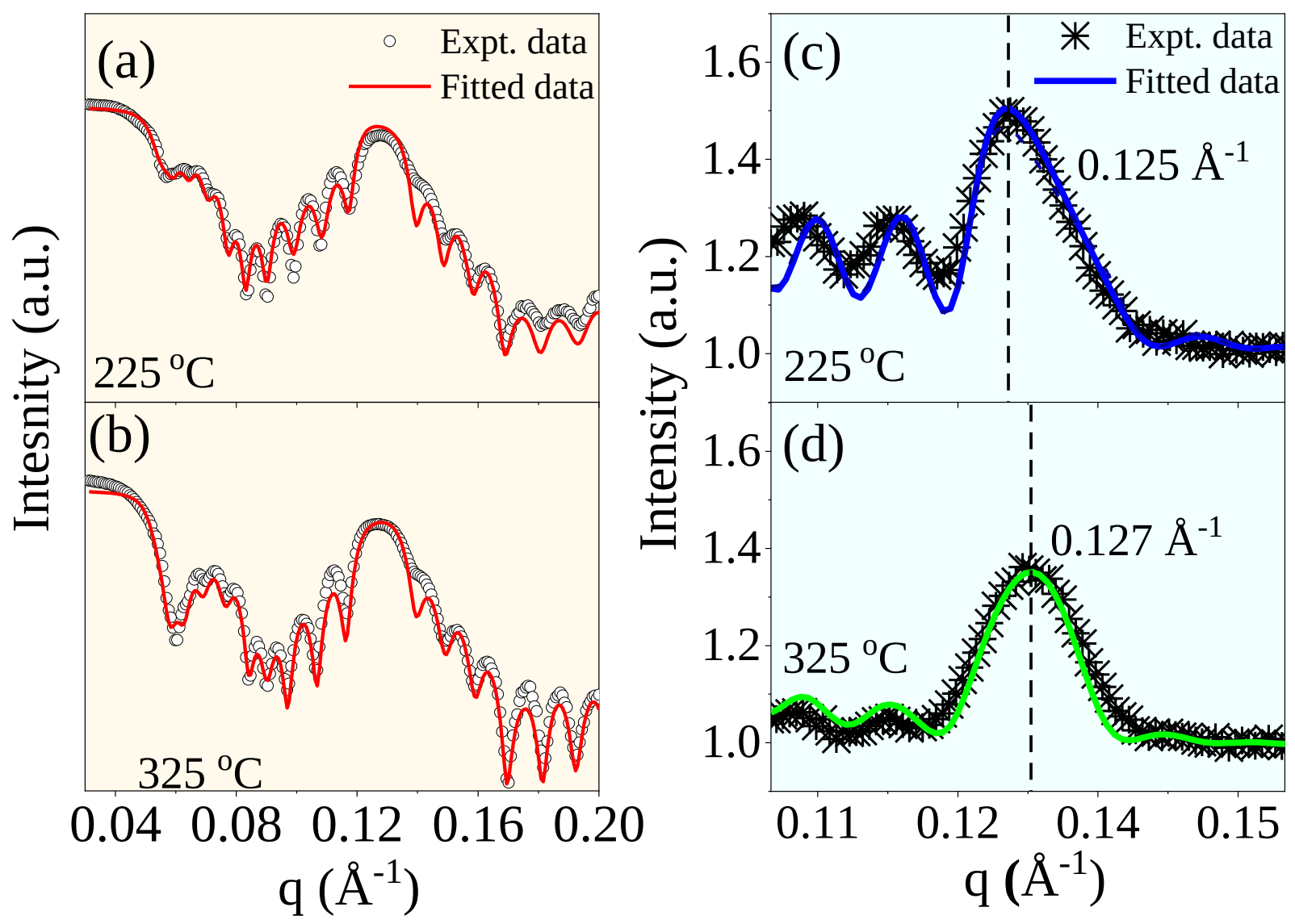

**Figure 6.** (a) and (b) show the reflectivity, (c) and (d) represent the normalized $^{57}$Fe fluorescence for the samples annealed at 225 °C and 325 °C. The scattered curve represents the experimental data, and the continuous curve represents the best fit.

To obtain qualitative information about the interfaces, the XRR and XRF data were fitted simultaneously. In multilayer systems, the reflectivity is strongly influenced by the Bragg response, and the fitting parameters are often highly correlated. Therefore, the simultaneous XRR/XRF fitting was focused on reproducing the main oscillatory features and extracting reliable trilayer structural parameters. Fittings of these patterns are plotted in the respective plots in continuous lines. After annealing at 225 °C, the $^{57}$Fe-on-Ag and Ag-on-$^{57}$Fe interfaces exhibit a modest increase in interfacial roughness, reaching 11.2 Å and 7.5 Å, respectively. This indicates that annealing promotes interfacial intermixing at both interfaces, thereby enhancing the inherent asymmetry in their structural profiles.

Upon further annealing at 325 °C, the effective interfacial roughness becomes comparable to the nominal Fe layer thickness. Consequently, the reflectivity data were modeled by treating the Fe layer as fully intermixed with the adjacent Ag layers. A similar study reported by G. Sharma et al. also revealed interface asymmetry in the as-prepared state [10]. It reports a larger roughness at the Fe-on-Ag interface than at the Ag-on-Fe interface. However, the annealing-induced evolution at intermediate temperatures differs from that of the present work. Such discrepancies may arise from differences in the initial microstructure due to variations in growth conditions and in the underlying standing-wave mirror structures (W/Si in the present work and $W/B_4C$ in G. Sharma et al.). Since interface diffusion and relaxation in interfaces are highly sensitive to the initial morphology and defect structure, even subtle differences in the growth environment can alter the balance between interface smoothing and intermixing during annealing. The present results indicate that intermixing-induced interface broadening dominates the thermal evolution, whereas G. Sharma et al. report an initial interface relaxation process before significant interdiffusion [10].

**Table 1:** Interface roughness and magnetic parameters obtained from MOKE and GI-NRS measurements for the pristine and annealed samples at 225 °C and 325 °C.

| **Temperature** | **Interface Roughness** | | **Magnetic properties** | | | |
|---|---|---|---|---|---|---|
| | | | **MOKE** | | **GI-NRS** | |
| | | | | | **$^{57}$Fe-on-Ag Interface** | **Ag-on-$^{57}$Fe interface** |
| | **$^{57}$Fe-on-Ag Error ± 1.0 Å** | **Ag-on-$^{57}$Fe Error ± 1.0 Å** | **(Hc) Error ±1 Oe** | **(Mr/Ms) Error ± 0.05** | **Bhf (T) Error ± 1** | **Bhf (T) Error ± 1** |
| **Pristine** | 9.7 Å | 6.4 Å | 7.0 Oe | 0.80 | 32.5, 29.4, 23.7, ~ 0 | 32.5, 29.4, 23.7, ~ 0 |
| **225 °C** | 11.2 Å | 7.5 Å | 8.4 Oe | 0.80 | 31.4, 28.4, 17.8, 1.4 | 31.4, 28.4, 17.8, 1.4 |
| **325 °C** | *$^{57}$Fe diffused with Ag layers* | | *Hysteresis loop without Hc and Mr/Ms* | | ~ 0, 1.26, 1.68 | |

To further understand the magnetic properties in detail, with an aim to study magnetic order in the annealed state, MOKE, CEMS, and GI-NRS measurements are performed, and the results are plotted in Fig. 7 (a-f), respectively. These measurements reveal a pronounced evolution in the magnetic properties of the samples upon annealing. The MOKE hysteresis loops recorded after annealing are shown in Fig. 7(a) and (b). At 225 °C, the sample exhibits a slight increase in coercivity compared to the as-prepared state [Fig. 7(a)]; the Hc and the Mr/Ms are 8.4 ± 1 Oe and 0.8 ± 0.05, respectively. In contrast, following annealing at 325 °C, the MOKE loop becomes essentially nonhysteretic, displaying an almost linear dependence on the applied magnetic field. Moreover, the signal does not reach saturation even under an applied field of ±3000 Oe. Consistent with this conclusion, a drastic modification of the CEMS spectra is observed after annealing at 325 °C [Fig. 7(d)]. The sextet structure present in the as-prepared and moderately annealed samples at 225 °C collapses into a singlet. The sextet spectrum was fitted using multiple Bhf components in the $^{57}Fe$ layer, with Bhf = 28.94, 31.49, 1.49, and 17.82 T. In contrast, the spectrum after annealing at 325 °C is well described by a single peak with a reduced hyperfine field of Bhf ≈ 1.68 T.

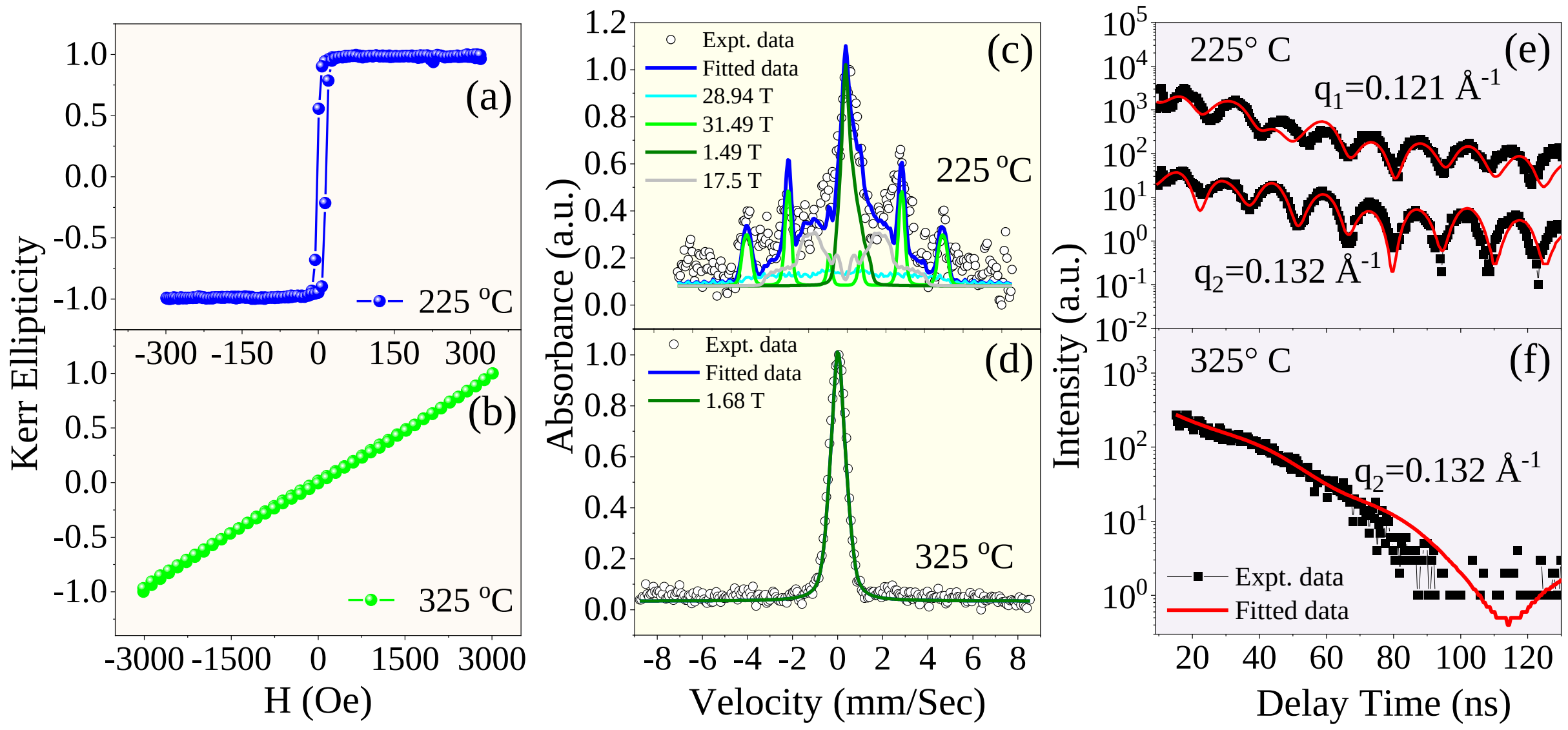

**Figure 7.** (a) and (b) show the MOKE hysteresis loops, (c) and (d) show the fitted CEMS spectra, (e) and (f) represent the fitted GINRS time spectra recorded at selected incident angles for samples annealed at 225 °C and 325 °C, respectively. The time spectra have been shifted vertically for clarity and the intensity is presented in log scale.

Unlike MOKE and CEMS, which probe magnetization averaged over the entire sampled volume, GI-NRS provides true depth selectivity with enhanced sensitivity through the formation of XSW. In view of this capability, GI-NRS measurements were performed for both annealed states under XSW conditions, and the corresponding data are shown in Fig. 7(e, f). For the sample annealed at 225 °C, GI-NRS measurements were carried out at 0.121 and 0.132 $Å^{-1}$, corresponding to standing-wave antinodes localized at the two interfaces, thereby enabling interface-selective probing of magnetic properties similar to the previous section. Both GI-NRS time spectra were fitted using the same set of Bhfs as obtained from the CEMS analysis for the annealed stage. However, the plots were well fitted by keeping different relative density distributions of the Bhf components at the two interfaces, as presented in the pie charts of Fig. 8. At 225 °C, the top interface retains a larger fraction of bulk-like $^{57}Fe$ environments, with the highest-field component (31.49 T) contributing 16%, compared to 6.6% at the bottom interface. The intermediate-field component at 28.49 T contributes comparably at the two interfaces (15.2% at the top and 14.6% at the bottom). In contrast, low-field components below 20 T (17.82 T and 1.49 T), associated with strongly diluted or superparamagnetic-like Fe environments, are more prominent at the bottom interface. Consequently, the top interface preserves a relatively higher proportion of bulk-like

Fe sites, indicative of a more ordered local environment and stronger magnetic coupling, whereas the bottom interface exhibits a broader Bhf distribution dominated by low-field contributions, establishing a pronounced magnetic asymmetry between the two interfaces.

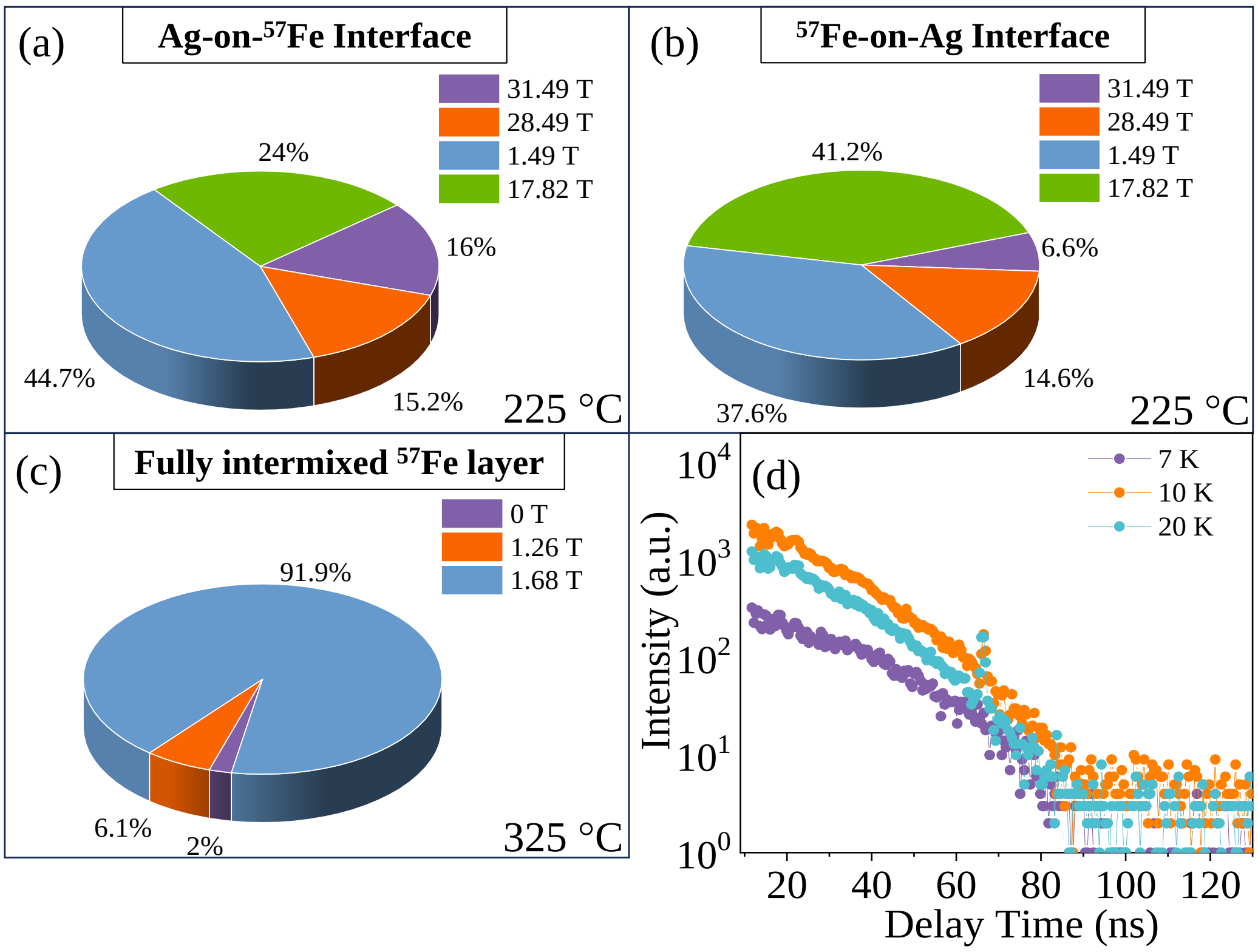

**Figure 8.** Percentage contribution of Bhfs in samples at (a) Ag-on-$^{57}$Fe and (b) $^{57}$Fe-on-Ag interface for 225 °C and (c) for the fully intermixed layer at 325 °C. (d) Shows the low-temperature NRS data for the 325 °C annealed sample for various temperatures.

In contrast, after annealing at 325 °C, the quantum beat patterns in the GI-NRS time spectra disappear [Fig. 7(f)], indicating the suppression of magnetic hyperfine splitting. The fitted spectra are well described by hyperfine field components of 1.68, 1.26, and ~ 0 T. The extracted Bhf distributions at different q values are found to be nearly identical across the depth of the $^{57}$Fe layer, and the corresponding density distribution is shown as a pie chart in Fig. 8(c) at one q only. It is clear that the Bhf distribution is dominated by very low-field and near-zero-field components. The 0 T Bhf component alone accounts for approximately 91.9 % of all $^{57}$Fe sites [30]. This indicates that most $^{57}$Fe atoms are in highly diluted or dynamically fluctuating magnetic environments, consistent with the loss of long-range ferromagnetic order. The interface roughness and magnetic parameters obtained from the MOKE and GI-NRS measurements for the pristine and annealed samples are summarized in Table 1.

After annealing at 325 °C, the MOKE hysteresis loop becomes linear with negligible coercivity and remanence, indicating the absence of stable ferromagnetic order and suggesting a superparamagnetic or paramagnetic state. This conclusion is further supported by the CEMS and NRS results. To probe the magnetic state in more detail, GI-NRS time spectra were recorded down to cryogenic temperatures. As shown in Fig. 8(d), the spectra measured at 20 K, 10 K, and 7 K exhibit similar decay profiles. Even at 7 K, no high-frequency quantum beats associated with static nuclear Zeeman splitting are observed. Instead, the spectra are dominated by a rapid, purely dynamical decay. The absence of long-lived oscillations or resolved magnetic hyperfine structure confirms the lack of magnetic blocking and indicates a highly diluted or dynamically fluctuating magnetic environment. The combined MOKE and CEMS measurements indicate the absence of stable ferromagnetic order after annealing at 325 °C. This behaviour may arise from the emergence of a superparamagnetic or paramagnetic state upon annealing.

### (iii) EXAFS measurements for probing local structure:

EXAFS measurements provide complementary information on the local coordination environment around Fe atoms and their evolution with temperature. The analysis of the EXAFS spectra provides the information about the local atomic structure for the selected atomic species, in our case the sample contains Fe atoms. To measure an absorption spectrum, one has to vary the energy of the incident beam. To keep the XSW field pattern constant, corrections of the incident angle were necessary. Usually, angular corrections are required to compensate for the varying penetration depths. The EXAFS spectra were recorded for both pristine and annealed samples in fluorescence mode at a grazing incidence angle of 0.124 $Å^{-1}$, corresponding to the maximum of the Fe fluorescence in the angular scan. In this geometry, the XSW antinode is positioned at the bottom interface, where intermixing effects are more pronounced. Several scans (> 4) were averaged to improve the signal-to-noise ratio. The incident X-ray intensity has been monitored during the data acquisition and used for the normalization of the fluorescence yield. The EXAFS data reduction was performed using Athena within Demeter Suite [38], including background subtraction, pre-edge, post-edge, and normalization. The reported bond distances are presented without phase-shift correction and therefore correspond to apparent radial positions in Fourier space. Figure 9 shows the EXAFS spectra taken for all the samples, where clear differences are observed in the amplitude and frequency of the EXAFS oscillations as a function of temperature. The Fourier transform (FT) of the EXAFS spectra reveals that the pristine sample exhibits features characteristic of bcc Fe, with a dominant first-shell peak corresponding to the Fe-Fe bond distance. At 225 °C, the EXAFS spectra exhibit a similar bcc Fe structure with a slight shift of the first shell peak towards a lower radial distance. At 325 °C, the EXAFS spectra change drastically, indicating a structural modification at the local level. A comparison of the experimental EXAFS spectra reveals that, while the main oscillatory features (at approximately 5, 6.5, and 8 $Å^{-1}$) remain at similar k positions for both the pristine and 225 °C samples, a clear reduction in amplitude is observed after annealing at 225 °C. The damping of these oscillations indicates an increase in the average structural disorder. In contrast, the spectrum of the 325 °C sample exhibits significant modifications. The dominant oscillatory features shift toward lower k values, and the relative intensity ratios differ markedly from those of the pristine and 225 °C samples. Notably, the most intense feature appears around 4.5 $Å^{-1}$, suggesting an enhanced contribution from a shorter interatomic distance or increased scattering from neighboring atoms. To quantify these structural changes, the k-weighted (χ(k)) spectra were fitted, and the best fit to the data are presented in Fig. 9. The quantitative results obtained from the analysis including coordination number (N), bond length (R), and Debye-Waller factor ($\sigma^2$) are summarized in Table 2.

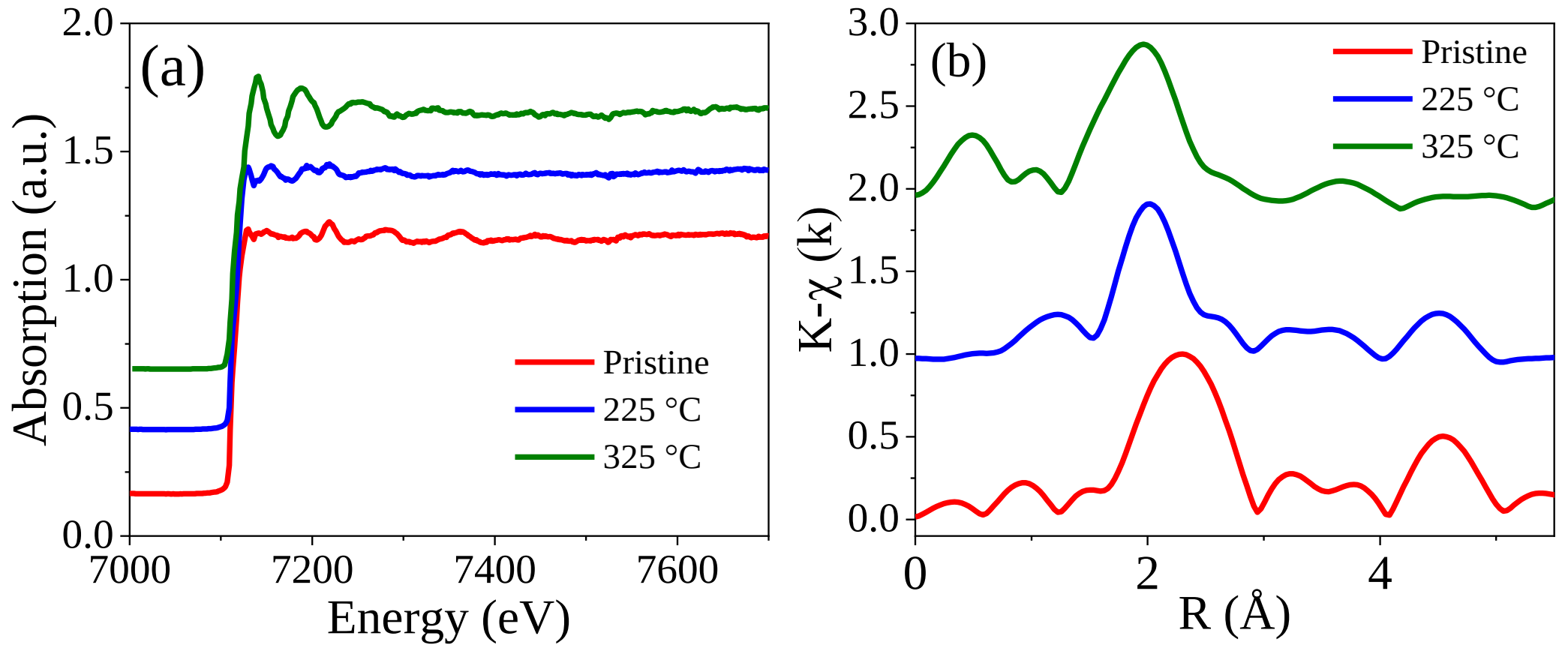

**Figure 9.** (a) shows the XAFS spectra for all the samples, and (b) shows the Fourier transform (FT) of the Fe k-edge EXAFS

The EXAFS spectra of the pristine sample are well reproduced using only Fe-Fe single-scattering paths of bcc Fe (Im-3m), indicating that the local structure probed by EXAFS remains predominantly bcc-like despite the presence of a thin intermixed interface formed during deposition [Fig. 10(a, d)]. The notations $R_1$, $R_2$, and $R_3$ represent the first, second, and third nearest neighbour (NN) bond distances, respectively, corresponding to the coordination shells of the bcc Fe lattice. For the pristine sample, the first-shell Fe-Fe bond distance was $R_1 = 2.48 \pm 0.01$ Å with a coordination number N = 8. The second and third-shell Fe-Fe distances were $R_2 = 2.82 \pm 0.01$ Å (N = 6) and $R_3 = 3.97 \pm 0.04$ Å (N = 12), respectively, further confirming the dominance of bulk-like bcc Fe coordination. The relatively small Debye-Waller factors ($\sigma^2 \approx 0.006$-$0.018$ $Å^2$) indicate a well-ordered Fe local structure. For the sample annealed at 225 °C, the EXAFS spectra are still adequately described by Fe-Fe single-scattering paths, indicating that the bcc-like Fe framework is largely preserved. However, subtle changes in the local coordination are observed, with a possible lattice relaxation (shrinkage) of the Fe-Fe shell radii to $R_1 = 2.44 \pm 0.02$ Å and $R_2 = 2.77 \pm 0.02$ Å. In addition, a third-shell Fe-Fe contribution appears at $R_3 \approx 3.95 \pm 0.04$ Å. These changes, together with the increased Debye-Waller factors ($\sigma^2 \approx 0.016$-$0.024$ $Å^2$), point to enhanced local disorder associated with thermally activated intermixing.

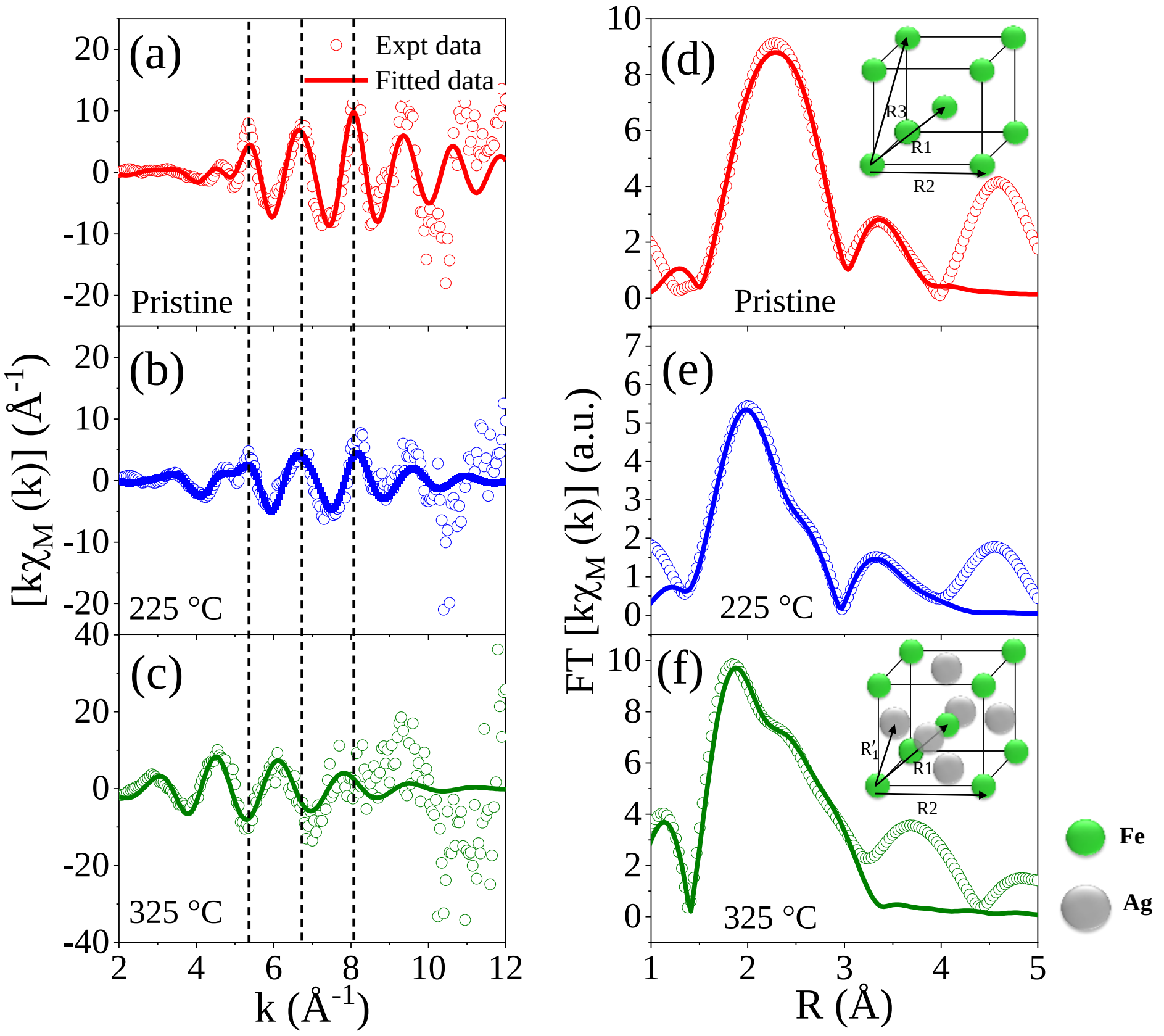

**Figure 10.** (a-c) shows the fitted $k^3$-weighted Fe k-edge EXAFS spectra for all the samples, and (d-f) shows the FT of the Fe k-edge EXAFS spectra for all the samples along with their best fit. The insets show the Fe unit cell with the 1st, 2nd, and 3rd shells considered for fitting.

**Table 2:** Results from the refinement of the EXAFS spectra measured in XSW conditions at Fe K-edge for all the samples. The fit quality (R-factor), $\Delta E_0$ (eV), $S_0^2$ are mentioned below.

| **Sample** | **Fit quality** | **Scattering paths** | | | | | | | | | | | |
|---|---|---|---|---|---|---|---|---|---|---|---|---|---|
| | | **Fe-Ag (1st shell)** **$\Delta E_0$= -3.217, $S_0^2$=0.75** | | | **Fe-Fe (1st shell)** **$\Delta E_0$= -3.217, $S_0^2$=0.75** | | | **Fe-Fe (2nd shell)** **$\Delta E_0$= -3.217, $S_0^2$=0.75** | | | **Fe-Fe (3rd shell)** **$\Delta E_0$= -3.217, $S_0^2$=0.75** | | |
| | **R-factor** | **N** | **$R'_1$ (Å)** | **$\sigma^2$ (Å$^2$)** | **N** | **$R_1$ (Å)** | **$\sigma^2$ (Å$^2$)** | **N** | **$R_2$ (Å)** | **$\sigma^2$ (Å$^2$)** | **N** | **$R_3$ (Å)** | **$\sigma^2$ (Å$^2$)** |
| **Pristine** | 0.01 | - | - | - | 8 | 2.48(1) | 0.017(1) | 6 | 2.82(1) | 0.006(1) | 12 | 3.97(4) | 0.018(3) |
| **225 °C** | 0.04 | - | - | - | 8 | 2.44(2) | 0.016(1) | 6 | 2.77(2) | 0.017(2) | 12 | 3.95(4) | 0.024(4) |
| **325 °C** | 0.03 | 5.0(8) | 1.86(1) | 0.018(3) | 8 | 2.63(2) | 0.016(3) | 6 | 3.12(2) | 0.009(2) | - | - | - |

In contrast, annealing at 325 °C leads to clear spectral modifications (Fig. 10(c, f)), including a reduced amplitude of the main Fourier-transform peak, reflecting enhanced disorder and changes in the Fe local coordination environment. In this case, the Debye-Waller factor cannot account for the modifications observed in the EXAFS signal. The experimental EXAFS data at the 325 °C annealed sample could no longer be adequately described using only the Fe-Fe scattering paths ($R_1$, $R_2$, $R_3$) of the bcc structure. To achieve a physically meaningful fit, the model was modified by incorporating Ag atoms into the local structure around Fe atoms and considering the Fe-Ag scattering path ($R'_1$), which improved the quality of the fit significantly. The Fe-Ag coordination number was allowed to vary during fitting, and the best fit points out to a value of $N \approx 5 \pm 0.87$, with a bond distance of $R'_1 \approx 1.86 \pm 0.01$ Å, suggesting the emergence of a modified local coordination environment involving Fe-Ag scattering contributions. At the same time, the Debye-Waller factors remain elevated, consistent with a highly disordered local environment. Despite the persistence of Fe-Fe correlations, the growing Fe-Ag coordination signifies progressive nanoscale intermixing. This locally intermixed and chemically disordered environment is consistent with the paramagnetic behaviour observed in the MOKE, CEMS, and NRS time spectra at 325 °C.

### (iv) Discussion:

The present work employs NRS under XSW conditions, which enables independent, depth-resolved characterization of both the $^{57}$Fe-on-Ag and Ag-on-$^{57}$Fe interfaces within a single trilayer by tuning the X-ray incidence angle [45,46]. This contrasts with conventional techniques, which either lack sufficient depth resolution or do not probe true interfaces. The present approach directly correlates local magnetic properties with interface-specific structure [13,29]. In Ag/$^{57}$Fe/Ag trilayer, the bottom $^{57}$Fe-on-Ag interface is significantly more diffuse than the top Ag-on-$^{57}$Fe interface. Such asymmetry is characteristic of multilayers composed of materials with dissimilar surface free energies [10] and can be traced to growth-induced segregation and interfacial intermixing [48]. Discontinuous Ag growth and the lower surface free energy of Ag favour Ag segregation during $^{57}$Fe deposition, leading to chemically intermixed and roughened $^{57}$Fe-on-Ag interfaces. In contrast, Ag deposited on $^{57}$Fe wets the surface more uniformly, producing a sharper and smoother Ag-on-$^{57}$Fe interface. This intrinsic structural asymmetry directly manifests in the magnetic response. The diffused $^{57}$Fe-on-Ag interface is characterized by a broad distribution of reduced hyperfine fields, whereas the sharper Ag-on-$^{57}$Fe interface retains a larger fraction of bulk-like $^{57}$Fe environments. The modest suppression of the bulk-like hyperfine field relative to α-Fe reflects residual disorder and weak Ag dilution. Strongly reduced hyperfine fields at the interfaces indicate substantial Fe-Ag chemical intermixing, as Ag nearest neighbours significantly suppress the Fe hyperfine field [49]. Nevertheless, the persistence of predominantly bcc-like Fe coordination in the pristine state indicates that this intermixing remains confined to a narrow interfacial region and does not disrupt the interior of the Fe layer.

Thermal annealing drives a progressive redistribution from bulk-like ferromagnetic Fe environments toward low- and near-zero-field components, signalling the breakdown of the continuous ferromagnetic layer. This behaviour reflects thermally activated Fe-Ag interdiffusion and phase segregation, which promote magnetic dilution and finite-size effects [50,51]. At 325 °C, the disappearance of hysteresis and the emergence of a linear magnetic response indicate the loss of long-range ferromagnetic order. Similar annealing-induced magnetic degradation has been reported in Fe-Ag granular systems, where chemical dilution and reduced coordination strongly suppress exchange interactions [52–54]. At this stage, three magnetic scenarios may be considered: formation of a continuous Fe-Ag alloy, superparamagnetic Fe clusters, or a paramagnetic state consisting of isolated Fe atoms or ultra-small clusters. Alloy formation can be excluded because Fe-Ag is thermodynamically immiscible, and even metastable Fe-rich alloys remain ferromagnetic and would exhibit hysteretic behaviour [52,53]. Superparamagnetism is also unlikely, as no magnetic blocking is observed down to low temperatures, implying negligibly small anisotropy barriers. The absence of magnetic blocking down to the lowest measured temperature (TB < 7 K) indicates that the magnetic entities present in the sample possess extremely small anisotropy barriers. Consequently, the observed paramagnetic component is attributed sto atomically dispersed $^{57}$Fe atoms and/or ultra-small $^{57}$Fe clusters embedded in an Ag-rich environment. In such systems, rapid spin fluctuations suppress the formation of static magnetic order on the Mössbauer timescale, resulting in the absence of resolved magnetic hyperfine splitting. This interpretation is consistent with both the Mössbauer spectra and the purely dynamical magnetic response observed in the nuclear resonant scattering measurements [54].

Overall, the results establish a direct correlation between interfacial morphology, chemical disorder, and magnetic phase stability in Ag/$^{57}$Fe/Ag trilayers. Annealing-induced interdiffusion progressively reduces Fe-Fe coordination and weakens exchange interactions, driving the system from an interface-modulated ferromagnetic state toward a paramagnetic regime. These findings underscore the decisive role of interface-specific roughness and chemical intermixing in governing magnetism in metallic heterostructures and provide a microscopic framework for engineering magnetic dilution through controlled thermal processing.

## Conclusion

This work demonstrates that X-ray standing waves generated using periodic multilayers can clearly resolve the two interfaces of a few nm thick $^{57}$Fe layer in Ag/$^{57}$Fe/Ag trilayers. This work reveals how intrinsic interfacial asymmetry and thermally driven intermixing govern the magnetic phase stability of Ag/$^{57}$Fe/Ag trilayers. The two chemically identical but structurally inequivalent interfaces exhibit markedly different magnetic environments in the pristine state, demonstrating that interface morphology alone can decisively influence local magnetism in metal/metal heterostructures. Controlled thermal annealing activates Fe-Ag interdiffusion, progressively reducing Fe-Fe coordination and weakening exchange interactions, which ultimately destroys long-range ferromagnetic order and drives the system into a paramagnetic regime at elevated temperatures. Resolving the magnetic contributions of individual interfaces within a single trilayer under XSW conditions provides direct microscopic evidence linking interface roughness, chemical disorder, and magnetic degradation. These insights establish a general framework for understanding and controlling interfacial magnetism in metallic multilayers and offer clear guidelines for the design of thermally robust spintronic heterostructures where magnetic properties are engineered through interface-specific control.

# Acknowledgment

Financial support from the Department of Science and Technology, Government of India (project CRG/2021/003094) and travel support within the framework of the India@DESY collaboration is gratefully acknowledged. Department of Science and Technology, New Delhi, for the fellowship in the form of DST INSPIRE (IF210723) is gratefully acknowledged. Thanks to Mr. Anil Gome for the XRF measurements.